\documentclass[12pt]{article}
 \usepackage{graphicx}
 \usepackage[cp1251]{inputenc}

\begin{document}
 \noindent {\footnotesize\it
   Astronomy Letters, 2022, Vol. 48, Issue 4, pp. 243-255}
 \newcommand{\dif}{\textrm{d}}

 \noindent
 \begin{tabular}{llllllllllllllllllllllllllllllllllllllllllllll}
 & & & & & & & & & & & & & & & & & & & & & & & & & & & & & & & & & & & & & &\\\hline\hline
 \end{tabular}

  \vskip 0.5cm
  \bigskip
\centerline{\bf\Large Kinematics of OB Stars with Data from }
 \centerline{\bf\Large the LAMOST and Gaia Catalogues}

 \bigskip
 \bigskip
 \centerline{\bf
            V. V. Bobylev\footnote[1]{e-mail: vbobylev@gaoran.ru} (1),
            A. T. Bajkova (1),
            G. M. Karelin (1)
            }
 \bigskip
 \centerline{\small\it(1)
 Pulkovo Astronomical Observatory of the Russian Academy of Sciences, St. Petersburg, Russia}

 \bigskip
 \bigskip
{\bf Abstract}---We have analyzed the kinematics of OB stars from the list by Xiang et al. (2021) that
contains $\sim$13 000 single OB stars. For these stars there are photometric distance estimates and proper motions from the Gaia catalogue and line-of-sight velocities from the LAMOST catalogue. Based on
a sample of single OB stars and using the photometric distances and proper motions of stars from the
Gaia EDR3 catalogue, we have found the group velocity components $(U_\odot,V_\odot,W_\odot)=(9.63,9.93,7.45)\pm(0.27,0.34,0.10)$~km s$^{-1}$, and the following parameters of the angular velocity of Galactic rotation:
 $\Omega_0=29.20\pm0.18$~km s$^{-1}$ kpc$^{-1}$,
 $\Omega^{'}_0=-4.150\pm0.046$~km s$^{-1}$ kpc$^{-2}$ and
 $\Omega^{''}_0=0.795\pm0.018$~km s$^{-1}$ kpc$^{-3}$,
where the error per unit weight $\sigma_0$ is 9.56~km s$^{-1}$ and $V_0=236.5\pm3.3$~km s$^{-1}$ (for the adopted $R_0=8.1\pm0.1$~kpc). Based on the same OB stars, we have found the residual velocity dispersions $(\sigma_1,\sigma_2,\sigma_3)=(15.13,9.69,7.98)\pm(0.07,0.05,0.04)$~km s$^{-1}$. We show that using the line-of-sight velocities increases
significantly the space velocity dispersion and leads to a biased estimate of the velocity $U_\odot$.  A comparison of the distances scales used has shown that the photometric distances from Xiang et al. (2021) should be lengthened approximately by 10\%.

\bigskip\noindent
{\bf DOI:} 10.1134/S1063773722040016


\newpage
\section*{INTRODUCTION}
Stars of spectral types O and B are young (a few Myr) massive (more than 10$M_\odot$) high-luminosity
stars. Owing to these properties, they are of great importance for studying the structure and kinematics of the Galaxy.

Various approaches are known for the spectral classification of O and B stars (Reed 1995; Zaal et al. 2001; Sota et al. 2011; Skiff 2014; Maiz Apell\'aniz et al. 2004, 2016), which is important for estimating their luminosities and, in the long run, photometric distances. When using photometric
measurements in the visual range, the accuracy of these distances is moderately high. For example,
in Reed (1995) the $1\sigma$ relative photometric distance error for single OB stars was estimated to be $\pm$ 39\%.

Using space photometric measurements from the Hipparcos (1997) satellite allowed the photometric
distances to 1\,207 OB stars to be determined with errors of $\pm$14\% (Wegner 2000). Applying up-to-date calibrations and photometric data in the infrared allows the photometric distances already to $\sim$15\,000 OB stars to be determined with errors of $\pm$12\% (Xiang et al. 2021).

The results of a kinematic analysis of stars depend strongly on the quality of their measured kinematic
characteristics. The accuracies of the proper motions of stars improve continuously; at present, they
have been measured for a large number of stars and, in particular, OB stars. The photometric distances
of OB stars were commonly used to analyze their spatial distribution and kinematics. The situation
has changed quite recently with the publication of more reliable stellar trigonometric parallaxes measured as a result of the Gaia space experiment (Prusti et al. 2016).

Stars of spectral types O and B were used by Oort (1927) to prove Lindblad’s hypothesis about
the Galactic rotation. Since then such stars have been repeatedly used to refine the Galactic rotation
parameters (Byl and Ovenden 1978; Miyamoto and Zhu 1998; Uemura et al. 2000; Dambis et al. 2001;
Branham 2002, 2006; Zabolotskikh et al. 2002; Popova and Loktin 2005; Zhu 2006; Mel’nik and
Dambis 2009, 2017; Gontcharov 2012; Bobylev and Bajkova 2018, 2019). Only their proper motions are
often taken for an analysis, because the line-of-sight velocities of single OB stars are measured with large errors.

OB stars are also used to study the structure and kinematics of the solar neighborhood, where young
open star clusters (Piskunov et al. 2006), OB associations (de Zeeuw et al. 1999; Dambis et al. 2001;
Mel’nik and Dambis 2020), the Gould Belt (Frogel and Stothers 1977; Torra et al. 2000), and the Local
Arm (Xu et al. 2021) are located. Such stars are of interest as tracers of the Galactic spiral structure (Y.M. Georgelin and Y.P. Georgelin 1976; Fern\'andez et al. 2001; Russeil 2003; Chen et al. 2019; Xu
et al. 2018, 2021).

A sample of OB stars with original photometric distance estimates, line-of-sight velocities from
the LAMOST (the Large Sky Area Multi-Object Fiber Spectroscopic Telescope, Cui et al. 2012) catalogue,
and proper motions and trigonometric parallaxes from the Gaia DR2 catalogue is described in Xiang et al. (2021). The goal of this paper is to redetermine the Galactic rotation parameters based on these OB stars and to compare the photometric distance scale with the distance scale based on the Gaia trigonometric parallaxes.

 \section*{METHOD}\label{method}
We have three stellar velocity components from observations: the line-of-sight velocity $V_r$ and the
two tangential velocity components $V_l=4.74r\mu_l\cos b$ and $V_b=4.74r\mu_b$ along the Galactic longitude $l$ and latitude $b$, respectively. All three velocities are expressed in km s$^{-1}$. Here, 4.74 is the dimension coefficient and $r$ is the stellar heliocentric distance in kpc. The proper motion components $\mu_l\cos b$ and $\mu_b$ are expressed in mas yr$^{-1}$. The velocities $U,V,W$ directed along the rectangular Galactic coordinate axes are calculated via the components $V_r, V_l, V_b$:

 \begin{equation}
 \begin{array}{lll}
 U=V_r\cos l\cos b-V_l\sin l-V_b\cos l\sin b,\\
 V=V_r\sin l\cos b+V_l\cos l-V_b\sin l\sin b,\\
 W=V_r\sin b                +V_b\cos b,
 \label{UVW}
 \end{array}
 \end{equation}
where the velocity $U$ is directed from the Sun toward the Galactic center, $V$ is in the direction of Galactic rotation, and $W$ is directed to the north Galactic pole. We can find two velocities, $V_R$ directed radially away from the Galactic center and $V_{circ}$ orthogonal to it pointing in the direction of Galactic rotation, based on the following relations:
 \begin{equation}
 \begin{array}{lll}
  V_{circ}= U\sin \theta+(V_0+V)\cos \theta, \\
       V_R=-U\cos \theta+(V_0+V)\sin \theta,
 \label{VRVT}
 \end{array}
 \end{equation}
where the position angle $\theta$ obeys the relation $\tan\theta=y/(R_0-x)$, $x,y,z$ are the rectangular heliocentric coordinates of the star (the velocities $U,V,W$ are directed along the corresponding $x,y,z$ axes), and $V_0$ s the linear rotation velocity of the Galaxy at the solar distance $R_0$.

\subsection*{Galactic Rotation Parameters}
To determine the parameters of the Galactic rotation curve, we use the equations derived from Bottlinger’s formulas, in which the angular velocity $\Omega$ is expanded into a series to terms of the second order of smallness in $r/R_0$:
\begin{equation} \begin{array}{lll}
 V_r=-U_\odot\cos b\cos l-V_\odot\cos b\sin l\\
 -W_\odot\sin b+R_0(R-R_0)\sin l\cos b\Omega^\prime_0\\
 +0.5R_0(R-R_0)^2\sin l\cos b\Omega^{\prime\prime}_0,
 \label{EQ-1}
 \end{array} \end{equation}
\begin{equation} \begin{array}{lll}
 V_l= U_\odot\sin l-V_\odot\cos l-r\Omega_0\cos b\\
 +(R-R_0)(R_0\cos l-r\cos b)\Omega^\prime_0\\
 +0.5(R-R_0)^2(R_0\cos l-r\cos b)\Omega^{\prime\prime}_0,
 \label{EQ-2}
 \end{array}\end{equation}
\begin{equation} \begin{array}{lll}
 V_b=U_\odot\cos l\sin b + V_\odot\sin l \sin b\\
 -W_\odot\cos b-R_0(R-R_0)\sin l\sin b\Omega^\prime_0\\
    -0.5R_0(R-R_0)^2\sin l\sin b\Omega^{\prime\prime}_0,
 \label{EQ-3}
 \end{array} \end{equation}
where $R$ is the distance from the star to the Galactic rotation axis,
$R^2=r^2\cos^2 b-2R_0 r\cos b\cos l+R^2_0$. The velocities $(U,V,W)_\odot$ are the mean group velocity of the sample, reflect the peculiar motion of the Sun, and, therefore, are taken with the opposite sign; $\Omega_0$ is the angular velocity of Galactic rotation at the solar distance $R_0$, the parameters $\Omega^{\prime}_0$ and $\Omega^{\prime\prime}_0$ are the corresponding derivatives of the angular velocity, and $V_0=R_0\Omega_0$. In this paper $R_0$ is taken to be $8.1\pm0.1$ kpc, according to the review by Bobylev and Bajkova (2021), where it was derived as a weighted mean of a large number of present-day individual estimates.

Solving the conditional equations ~(\ref{EQ-1})--(\ref{EQ-3}) by the least-squares method (LSM), we can find six unknowns: $(U,V,W)_\odot,$ $\Omega_0$, $\Omega^{\prime}_0$, and $\Omega^{\prime\prime}_0$. In the LSM solution of only one conditional equation (\ref{EQ-1}) we can find
only five unknowns: $(U,V,W)_\odot,$ $\Omega^{\prime}_0$, and $\Omega^{\prime\prime}_0$.

The velocities $U,$ $V$, and $W$ in Eqs. (\ref{UVW}) and (\ref{VRVT}) were freed from the peculiar solar velocity $U_\odot,$ $V_\odot$, and $W_\odot$ with the values found through the LSM solution of the kinematic equations (\ref{EQ-1})--(\ref{EQ-3}).

 \subsection*{Residual Velocity Ellipsoid}\label{sigm-123}
We use the following well-known method (Ogorodnikov 1965) to estimate the stellar residual velocity
dispersions. We consider six second-order moments $a,b,c, f,e,$ and $d$:
\begin{equation}
 \begin{array}{lll}
 a=\langle U^2\rangle-\langle U^2_\odot\rangle,\\
 b=\langle V^2\rangle-\langle V^2_\odot\rangle,\\
 c=\langle W^2\rangle-\langle W^2_\odot\rangle,\\
 f=\langle VW\rangle-\langle V_\odot W_\odot\rangle,\\
 e=\langle WU\rangle-\langle W_\odot U_\odot\rangle,\\
 d=\langle UV\rangle-\langle U_\odot V_\odot\rangle,
 \label{moments}
 \end{array}
 \end{equation}
which are the coefficients of the surface equation
 \begin{equation}
 ax^2+by^2+cz^2+2fyz+2ezx+2dxy=1,
 \end{equation}
and the components of the symmetric residual velocity moment tensor
 \begin{equation}
 \left(\matrix {
  a& d & e\cr
  d& b & f\cr
  e& f & c\cr }\right).
 \label{ff-5}
 \end{equation}
The following six equations are used to determine the values of this tensor:
\begin{equation}
 \begin{array}{lll}
 V^2_l= a\sin^2 l+b\cos^2 l\sin^2 l\\
 -2d\sin l\cos l,
  \label{EQsigm-1}
  \end{array} \end{equation}
\begin{equation} \begin{array}{lll}
 V^2_b= a\sin^2 b\cos^2 l+b\sin^2 b\sin^2 l\\
 +c\cos^2 b
 -2f\cos b\sin b\sin l\\
 -2e\cos b\sin b\cos l\\
 +2d\sin l\cos l\sin^2 b,
 \label{EQsigm-2}
 \end{array}
 \end{equation}
\begin{equation}
 \begin{array}{lll}
 V_lV_b= a\sin l\cos l\sin b\\
 +b\sin l\cos l\sin b\\
 +f\cos l\cos b-e\sin l\cos b\\
 +d(\sin^2 l\sin b-\cos^2\sin b),
 \label{EQsigm-3}
 \end{array}
 \end{equation}
\begin{equation}
 \begin{array}{lll}
 V^2_r= a\cos^2 b\cos^2 l+b\cos^2 b\sin^2 l\\
 +c\sin^2 b
 +2f\cos b\sin b\sin l\\
 +2e\cos b\sin b\cos l\\
 +2d\sin l\cos l\cos^2 b,
 \label{EQsigm-4}
 \end{array}
 \end{equation}
\begin{equation}
 \begin{array}{lll}
 V_b V_r=-a\cos^2 l\cos b\sin b\\
 -b\sin^2 l\sin b\cos b+c\sin b\cos b\\
 +f(\cos^2 b\sin l-\sin l\sin^2 b)\\
 +e(\cos^2 b\cos l-\cos l\sin^2 b)\\
 -d(\cos l\sin l\sin b\cos b\\
 +\sin l\cos l\cos b\sin b),
 \label{EQsigm-5}
 \end{array}
 \end{equation}
\begin{equation}
 \begin{array}{lll}
 V_l V_r=-a\cos b\cos l\sin l\\
     +b\cos b\cos l\sin l\\
    +f\sin b\cos l-e\sin b\sin l\\
    +d(\cos b\cos^2 l-\cos b\sin^2 l),
 \label{EQsigm-6}
 \end{array}
 \end{equation}
which are solved by the least-squares method for the six unknowns $a,b,c,f,e,$ and $d$. The eigenvalues of the tensor (\ref{ff-5}) $\lambda_{1,2,3}$ are then found from the solution of the secular equation
 \begin{equation}
 \left|\matrix
 {
a-\lambda&          d&        e\cr
       d & b-\lambda &        f\cr
       e &          f&c-\lambda\cr
 }
 \right|=0.
 \label{ff-7}
 \end{equation}
The eigenvalues of this equation are equal to the reciprocals of the squares of the semiaxes of the velocity moment ellipsoid and, at the same time, the squares of the semiaxes of the residual velocity ellipsoid:
 \begin{equation}
 \begin{array}{lll}
 \lambda_1=\sigma^2_1, \lambda_2=\sigma^2_2, \lambda_3=\sigma^2_3,\\
 \lambda_1>\lambda_2>\lambda_3.
 \end{array}
 \end{equation}
The directions of the principal axes of the tensor (\ref{ff-7}) $L_{1,2,3}$ and $B_{1,2,3}$ are found from the relations
 \begin{equation}
 \tan L_{1,2,3}={{ef-(c-\lambda)d}\over {(b-\lambda)(c-\lambda)-f^2}},
 \label{ff-41}
 \end{equation}
 \begin{equation}
 \tan B_{1,2,3}={{(b-\lambda)e-df}\over{f^2-(b-\lambda)(c-\lambda)}}\cos L_{1,2,3}.
 \label{ff-42}
 \end{equation}

 \section*{DATA}
In this paper we use a sample of OB stars for which Xiang et al. (2021) determined the photometric
distances based on near-infrared data. The catalogue by Xiang et al. (2021) contains 16\,002 entries with data on the stars some of which were observed several times. The stars are provided with the proper motions from the Gaia DR2 catalogue (Brown et al. 2018) and the line-of-sight velocities from the LAMOST catalogue.

The LAMOST project is a deep (to $r<18.5^m$) spectroscopic sky survey in the optical wavelength
range ($\lambda: 3700-9000$~{\AA}) with a low spectral resolution ($R\sim1 800$). The observations are carried out at a Schmidt telescope with a 4-m mirror. According to Xiang et al. (2017), the errors in the line-of-sight velocities of late-type stars in the LAMOST catalogue are $\pm$5 km s$^{-1}$.

The spectroscopic classification of OB stars from the LAMOST\,DR5 catalogue was performed by Liu
et al. (2019). The list by Xiang et al. (2021) contains 15\,184 OB stars that were observed at least once. Among them there are 13\,029 stars marked as single ones. The line-of-sight velocities are available virtually for each star. In some cases, there is no information about the measurements of stellar proper motions. Figure 1 presents a histogram of random errors in the line-of-sight velocities of single OB stars. As can be seen from this figure, the random errors for the overwhelming majority of these stars is less than 5 km s$^{-1}$.

At present, the Gaia EDR3 (Gaia Early Data Release 3, Brown et al. 2021) version has been
published, where, in comparison with the previous Gaia DR2 version, the trigonometric parallaxes and
proper motions were improved approximately by 30\% for $\sim$1.5 billion stars. The trigonometric parallaxes for $\sim$500 million stars were measured with errors less than 0.2 mas. For stars with magnitudes $G<15^m$ the random measurement errors of the proper motions lie within the range 0.02–0.04 mas yr$^{-1}$, and they increase dramatically for fainter stars. On the whole, the proper motions for about a half of the stars in the catalogue were measured with a relative error less than 10\%.

We identified the list of OB stars by Xiang et al. (2021) with the Gaia\,EDR3 catalogue and found
14\,532 common stars. The distribution of OB stars from the catalogue by Xiang et al. (2021) in projection onto the Galactic $XY$ plane is shown in Fig.~\ref{f-10-XY}. To construct the figure, we took stars with line-of-sight velocity errors less than 10 km s$^{-1}$ and distance errors less than 10\%. The stars with the photometric distances and with the distances calculated via the Gaia\,EDR3 trigonometric parallaxes are shown. Note that the distribution of all 13\,000 OB stars on the $XY$ plane can be seen in Fig. 12 from Xiang et al. (2021). Unfortunately, no manifestations of the spiral structure are seen in both Fig.~\ref{f-10-XY}a and Fig.~\ref{f-10-XY}b.

The stars of the LAMOST program are observed from the northern hemisphere and, therefore, part of
the southern sky is unobservable. This explains the absence of stars in the fourth Galactic quadrant on
the Galactic $XY$ plane (Fig.~\ref{f-10-XY}).

In Fig.~\ref{f-distances} the distances to the OB stars calculated via the Gaia\,EDR2 and Gaia\,EDR3 trigonometric parallaxes are plotted against the photometric distances. To calculate the distances to the OB stars via the Gaia\,EDR2 and Gaia\,EDR3 trigonometric parallaxes, we used the individual corrections to the parallaxes for each star given by Xiang et al. (2021). We can see from Fig.~\ref{f-distances}a that the distance scales there are in satisfactory agreement up to distances no more than 1.5 kpc from the Sun.

There is much better agreement between the distances to the OB stars calculated via the Gaia\,EDR3
trigonometric parallaxes and the photometric distances (Fig.~\ref{f-distances}b). The discrepancy between these two distance scales becomes apparent at distances greater than $\sim$4 kpc. On the whole, we can conclude that the Gaia distance scale is longer than the photometric distance scale by Xiang et al. (2021), with the relation between these distance scale being nonlinear.

\begin{figure}[t]
{ \begin{center}
  \includegraphics[width=0.5\textwidth]{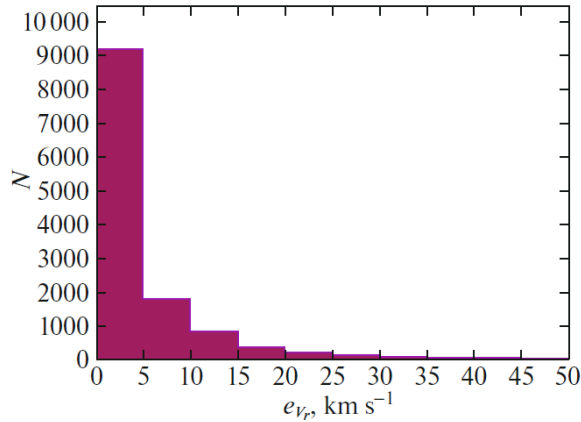}
  \caption{
Histogram of random errors in the line-of-sight velocities of single OB stars from the list by Xiang
et al. (2021).
  }
 \label{f-eRV}
\end{center}}
\end{figure}
\begin{figure}[t]
{ \begin{center}
  \includegraphics[width=0.85\textwidth]{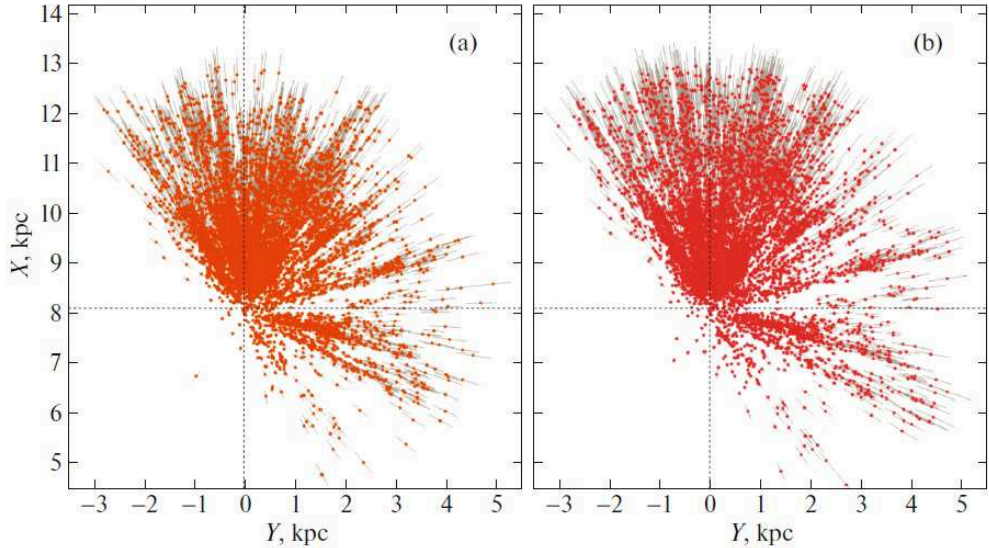}
  \caption{
Distribution of OB stars with the photometric distances (a) and OB stars with the distances calculated via the Gaia\,EDR3 trigonometric parallaxes (b) on the Galactic $XY$ plane.
  }
 \label{f-10-XY}
\end{center}}
\end{figure}
\begin{figure}[t]
{ \begin{center}
  \includegraphics[width=0.85\textwidth]{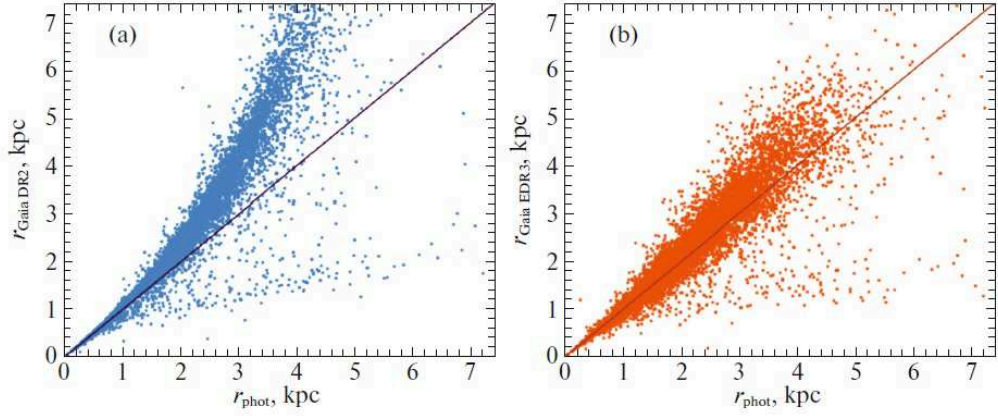}
  \caption{
Distances to the OB stars calculated via the Gaia\,DR2 trigonometric parallaxes versus photometric distances (a) and distances calculated via the Gaia\,EDR3 parallaxes versus photometric distances (b); the diagonal coincidence line is given on each panel.
  }
 \label{f-distances}
\end{center}}
\end{figure}
\begin{figure}[t]
{ \begin{center}
  \includegraphics[width=0.85\textwidth]{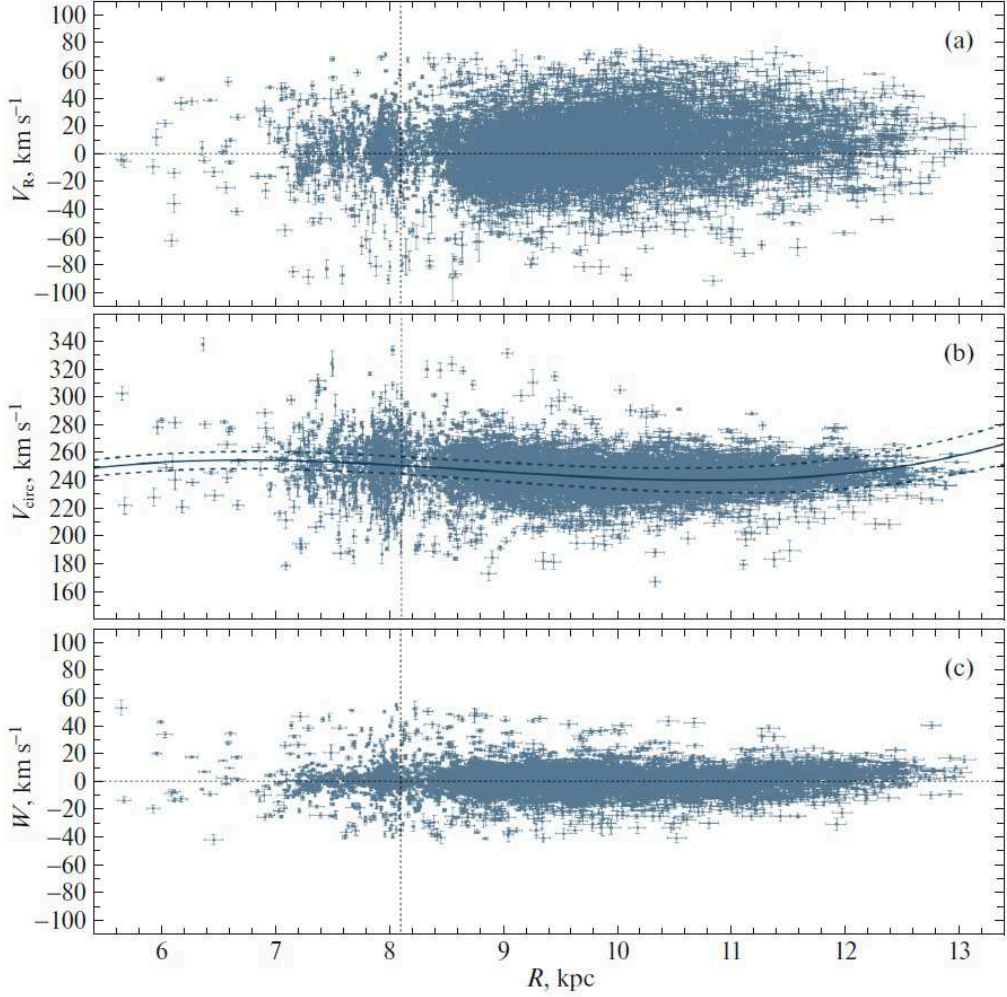}
  \caption{
(a) Radial velocities of the OB stars, $V_R$, (b) circular velocities $V_{circ}$ of the stars from this sample, and (c) their vertical velocities $W$ versus distance $R$; the vertical line marks the position of the Sun.
}
 \label{f-10-RT}
\end{center}}
\end{figure}
\begin{figure}[t]
{ \begin{center}
  \includegraphics[width=0.85\textwidth]{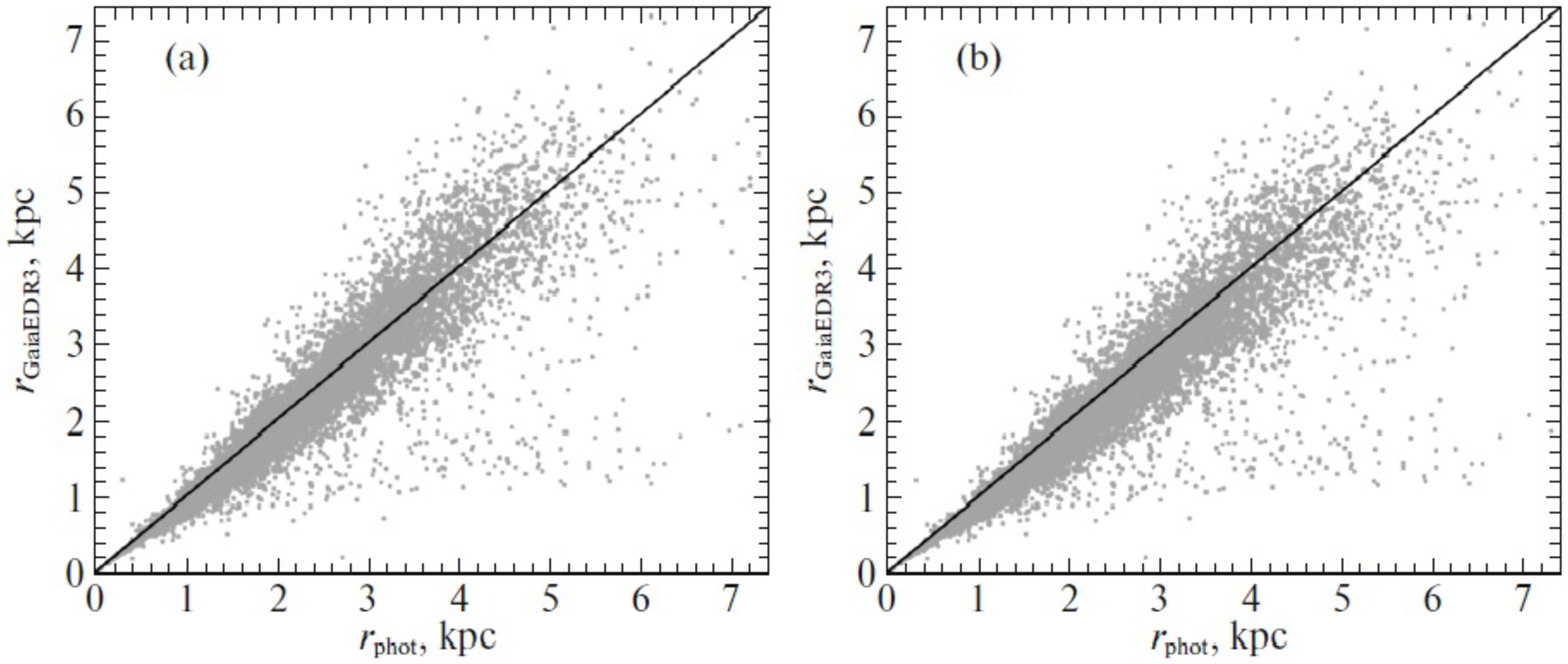}
  \caption{
Distances $r$ to the OB stars calculated via the Gaia\,EDR3 trigonometric parallaxes versus photometric ones increased by 10\% (a) and distances calculated via the Gaia\,EDR3 parallaxes versus photometric ones increased by 15\% (b); the diagonal coincidence line is presented on each panel.
  }
 \label{f-dist-2}
\end{center}}
\end{figure}

 \section*{RESULTS}
 \subsection*{Galactic Rotation Parameters}
It is well known that among the OB stars there are many runaway ones (see, e.g., Tetzlaff et al. 2011),
i.e., stars with high (more than 50 km s$^{-1}$) peculiar velocities. In addition, there are stars with large line-of-sight velocity errors that are not always rejected by the 3$\sigma$ criterion. To get rid of great outliers, we use the following constraints on the space velocities:
 \begin{equation}
 \begin{array}{rcl}
       |U|<80~\hbox{km s$^{-1}$},\\
       |V|<80~\hbox{km s$^{-1}$},\\
       |W|<50~\hbox{km s$^{-1}$}.
  \label{cut}
 \end{array}
 \end{equation}
The velocities $U,$ $V$ and $W$ here are the residual ones, since the peculiar solar motion and the Galactic rotation curve were subtracted from them.

The model parameters found from the OB stars using the photometric distances are presented in Table
~\ref{t:01}. The stellar proper motions from the Gaia\,DR2 catalogue and the line-of-sight velocities from the LAMOST catalogue were used here. The kinematic equations were solved by three methods: using all data, based only on the line-of-sight velocities, and based only on the stellar proper motions. A fairly smooth behavior of all parameters in all columns can be noted, i.e., the parameters found depend weakly on the distance error. The main feature is that (i) we have a very large error per unit weight $\sigma_0$ deduced from the line-of-sight velocities, (ii) the velocities $U_\odot$ found from the line-of-sight velocities differ noticeably from those found from the proper motions of the OB stars, and (iii) the parameters found by analyzing the proper motions are most credible, since they were determined with the smallest error per unit weight.

The error per unit weight $\sigma_0$ is determined when solving the conditional equations ~(\ref{EQ-1})--(\ref{EQ-3}) as a mean the residuals. This quantity characterizes the stellar residual velocity dispersion averaged over three directions. It is well known that the residual velocity dispersion of HI clouds in the Galactic disk is about 5 km s$^{-1}$, while the velocity dispersion of OB stars lies within the range 8–10 km s$^{-1}$.

Figure~\ref{f-10-RT} presents the radial, $V_R$, circular, $V_{circ}$, and vertical, $W$, velocities of the OB stars. The Galactic rotation curve shown in Fig.~\ref{f-10-RT}b was constructed with the parameters given in the first column of Table~\ref{t:01}. No influence of the spiral density wave that might be expected when analyzing young stars is seen in both Fig.~\ref{f-10-XY} and Fig.~\ref{f-10-RT}. Therefore, in this paper we did not include the terms describing the perturbations from the Galactic spiral density wave in the kinematic model.

 \begin{table}[p] \caption[]{\small
The kinematic parameters found from the OB stars based on all three conditional equations ~(\ref{EQ-1})--(\ref{EQ-3}) are given in the upper part of the table; those derived only from the LAMOST line-of-sight velocities based on Eq.~(\ref{EQ-1}) are given in the middle part of the table; those derived only from the Gaia\,DR2 proper motions based on the two Eqs.~(\ref{EQ-2}) and (\ref{EQ-3}) are given in the lower part of the table; the distances to the stars are photometric; $N_\star$ is the number of stars used, ${\overline r}$ is the mean radius of the sample.
 }
  \begin{center}  \label{t:01}
  \small
  \begin{tabular}{|l|r|r|r|r|r|}\hline
  Parameters  & $\sigma_r/r<10\%$ &
           $\sigma_r/r<15\%$ & $\sigma_r/r<20\%$ & $\sigma_r/r<30\%$ \\\hline

                 $N_\star$       & 9756 & 11738 & 11900 & 11943\\
        ${\overline r},$ kpc     & 1.95 &  2.20 &  2.22 &  2.22\\
  &&&&\\
   $U_\odot,$ km s$^{-1}$ & $13.47\pm0.17$ & $13.79\pm0.19$ & $13.75\pm0.16$ & $13.68\pm0.16$\\
   $V_\odot,$ km s$^{-1}$ & $10.91\pm0.29$ & $11.48\pm0.30$ & $11.45\pm0.29$ & $11.47\pm0.29$\\
   $W_\odot,$ km s$^{-1}$ & $ 6.94\pm0.15$ & $ 6.51\pm0.15$ & $ 6.49\pm0.15$ & $ 6.45\pm0.14$\\

   $\Omega_0,$ km s$^{-1}$ kpc$^{-1}$
      &$ 30.85\pm0.17$ & $ 30.75\pm0.16$ & $30.71\pm0.16$ & $ 30.66\pm0.16$\\
  $\Omega^{'}_0,$ km s$^{-1}$ kpc$^{-2}$
      &$-4.350\pm0.042$& $-4.193\pm0.038$&$-4.187\pm0.038$& $-4.173\pm0.037$\\
 $\Omega^{''}_0,$ km s$^{-1}$ kpc$^{-3}$
      &$ 0.852\pm0.019$& $ 0.757\pm0.015$& $0.755\pm0.015$& $ 0.753\pm0.014$\\
   $\sigma_0,$ km s$^{-1}$ &       $14.77$ &       $15.28$ &       $15.38$ &       $15.25$\\
        $V_0,$ km s$^{-1}$ & $249.9\pm3.4$ & $249.1\pm3.3$ & $248.7\pm3.3$ & $248.4\pm3.3$\\
 \hline

   $U_\odot,$ km s$^{-1}$ & $14.39\pm0.28$ & $14.84\pm0.26$ & $14.83\pm0.25$ & $14.81\pm0.25$\\
   $V_\odot,$ km s$^{-1}$ & $11.41\pm0.56$ & $11.51\pm0.55$ & $11.44\pm0.55$ & $11.46\pm0.55$\\

  $\Omega^{'}_0,$ km s$^{-1}$ kpc$^{-2}$
      &$-4.847\pm0.099$& $-4.605\pm0.090$&$-4.601\pm0.084$& $-4.587\pm0.090$\\
 $\Omega^{''}_0,$ km s$^{-1}$ kpc$^{-3}$
      &$ 1.352\pm0.072$& $ 1.127\pm0.057$& $1.121\pm0.043$& $ 1.108\pm0.056$\\
   $\sigma_0,$ km s$^{-1}$ &    $21.94$ &    $22.44$ &   $22.52$ &   $22.54$\\

 \hline

   $U_\odot,$ km s$^{-1}$ & $9.53\pm0.27$ & $ 9.15\pm0.25$ & $ 9.09\pm0.25$ & $ 9.13\pm0.25$\\
   $V_\odot,$ km s$^{-1}$ & $9.95\pm0.34$ & $10.95\pm0.32$ & $11.01\pm0.32$ & $10.98\pm0.32$\\
   $W_\odot,$ km s$^{-1}$ & $7.03\pm0.10$ & $ 6.63\pm0.10$ & $ 6.57\pm0.10$ & $ 6.55\pm0.10$\\

   $\Omega_0,$ km s$^{-1}$ kpc$^{-1}$
      &$ 29.06\pm0.18$ & $ 28.70\pm0.17$ & $28.62\pm0.17$ & $ 29.65\pm0.17$\\
  $\Omega^{'}_0,$ km s$^{-1}$ kpc$^{-2}$
      &$-4.167\pm0.046$& $-3.967\pm0.041$&$-3.943\pm0.040$& $-3.951\pm0.040$\\
 $\Omega^{''}_0,$ km s$^{-1}$ kpc$^{-3}$
      &$ 0.814\pm0.018$& $ 0.722\pm0.014$& $0.713\pm0.014$& $ 0.716\pm0.014$\\

   $\sigma_0,$ km s$^{-1}$ &       $9.65$ &       $10.01$ &       $10.07$ &       $10.09$\\
        $V_0,$ km s$^{-1}$ & $235.4\pm3.3$ & $232.5\pm3.2$ & $231.8\pm3.2$ & $232.1\pm3.2$\\

 $(\Omega^{'}_0)_{V_l}/(\Omega^{'}_0)_{V_r}$
     & $0.86\pm0.01$ &  $0.86\pm0.01$ & $0.86\pm0.01$ &  $0.86\pm0.01$  \\

 \hline
\end{tabular}\end{center} \end{table}

 \begin{table}[p] \caption[]{\small
The kinematic parameters found from the OB stars based on all three conditional equations ~(\ref{EQ-1})--(\ref{EQ-3}) are given in the upper part of the table; those derived only from the LAMOST line-of-sight velocities based on Eq.~(\ref{EQ-1}) are given in the middle part of the table; those derived only from the Gaia\,DR2 proper motions based on the two Eqs. (\ref{EQ-2}) and (\ref{EQ-3}) are given in the lower part of the table; the distances to the stars were calculated based on the Gaia\,DR2 parallaxes; $N_\star$ is the number of stars used, ${\overline r}$ is the mean radius of the sample.
 }
  \begin{center}  \label{t:02}
  \small
  \begin{tabular}{|l|r|r|r|r|r|}\hline
  Parameters  & $\sigma_\pi/\pi<10\%$ &
           $\sigma_\pi/\pi<15\%$ & $\sigma_\pi/\pi<20\%$ & $\sigma_\pi/\pi<30\%$ \\\hline

                 $N_\star$       & 6979 & 9352 & 10356 & 10750\\
        ${\overline r},$ kpc     & 1.84 & 2.24 &  2.42 &  2.48\\
  &&&&\\
   $U_\odot,$ km s$^{-1}$ & $11.85\pm0.22$ & $13.11\pm0.19$ & $13.27\pm0.18$ & $13.26\pm0.18$\\
   $V_\odot,$ km s$^{-1}$ & $10.60\pm0.33$ & $10.95\pm0.31$ & $11.42\pm0.32$ & $11.38\pm0.31$\\
   $W_\odot,$ km s$^{-1}$ & $ 7.86\pm0.19$ & $ 8.28\pm0.17$ & $ 8.18\pm0.16$ & $ 8.17\pm0.16$\\

   $\Omega_0,$ km s$^{-1}$ kpc$^{-1}$
      &$ 28.65\pm0.19$ & $ 28.76\pm0.16$ & $28.51\pm0.15$ & $ 28.43\pm0.15$\\
  $\Omega^{'}_0,$ km s$^{-1}$ kpc$^{-2}$
      &$-4.019\pm0.049$& $-3.918\pm0.037$&$-3.839\pm0.035$& $-3.825\pm0.034$\\
 $\Omega^{''}_0,$ km s$^{-1}$ kpc$^{-3}$
      &$ 0.688\pm0.023$& $ 0.625\pm0.013$& $0.601\pm0.011$& $ 0.594\pm0.011$\\
   $\sigma_0,$ km s$^{-1}$ &       $15.34$ &       $15.92$ &       $16.06$ &       $16.10$\\
        $V_0,$ km s$^{-1}$ & $232.1\pm3.8$ & $232.9\pm3.2$ & $230.9\pm3.1$ & $230.3\pm3.1$\\
 \hline

   $U_\odot,$ km s$^{-1}$ & $12.49\pm0.33$ & $13.79\pm0.29$ & $14.24\pm0.27$ & $14.34\pm0.27$\\
   $V_\odot,$ km s$^{-1}$ & $11.07\pm0.59$ & $10.78\pm0.57$ & $11.03\pm0.57$ & $11.10\pm0.57$\\

  $\Omega^{'}_0,$ km s$^{-1}$ kpc$^{-2}$
      &$-4.342\pm0.111$& $-4.167\pm0.088$&$-4.136\pm0.084$& $-4.110\pm0.083$\\
 $\Omega^{''}_0,$ km s$^{-1}$ kpc$^{-3}$
      &$ 1.237\pm0.077$& $ 1.046\pm0.048$& $1.005\pm0.043$& $ 0.980\pm0.042$\\
   $\sigma_0,$ km s$^{-1}$ &       $21.43$ &       $21.82$ &       $22.07$ &       $22.20$\\

 \hline

   $U_\odot,$ km s$^{-1}$ & $9.95\pm0.33$ & $10.50\pm0.31$ & $10.46\pm0.18$ & $10.43\pm0.30$\\
   $V_\odot,$ km s$^{-1}$ & $9.30\pm0.41$ & $11.07\pm0.35$ & $11.36\pm0.34$ & $11.31\pm0.34$\\
   $W_\odot,$ km s$^{-1}$ & $7.96\pm0.14$ & $ 8.37\pm0.12$ & $ 8.29\pm0.12$ & $ 8.25\pm0.12$\\

   $\Omega_0,$ km s$^{-1}$ kpc$^{-1}$
      &$ 29.14\pm0.20$ & $ 29.47\pm0.17$ & $29.23\pm0.17$ & $ 29.28\pm0.17$\\
  $\Omega^{'}_0,$ km s$^{-1}$ kpc$^{-2}$
      &$-4.220\pm0.056$& $-4.050\pm0.040$&$-3.988\pm0.037$& $-3.990\pm0.036$\\
 $\Omega^{''}_0,$ km s$^{-1}$ kpc$^{-3}$
      &$ 0.719\pm0.023$& $ 0.637\pm0.012$& $0.623\pm0.011$& $ 0.620\pm0.010$\\
   $\sigma_0,$ km s$^{-1}$ &       $10.96$ &       $11.40$ &       $11.50$ &       $11.55$\\
        $V_0,$ km s$^{-1}$ & $236.1\pm3.3$ & $238.7\pm3.3$ & $236.7\pm3.2$ & $237.1\pm3.2$\\

 $(\Omega^{'}_0)_{V_l}/(\Omega^{'}_0)_{V_r}$
     & $0.97\pm0.01$ &  $0.97\pm0.01$ & $0.96\pm0.01$ &  $0.97\pm0.01$  \\

 \hline
\end{tabular}\end{center} \end{table}

The model parameters found from the OB stars using data from the Gaia\,DR2 and Gaia\,EDR3 catalogues
are presented in Table~\ref{t:02} and in Tables~\ref{t:03} and \ref{t:04}, respectively. The results were obtained with constraints either on the distance errors $\sigma_r/r$ or on the parallax errors $\sigma_\pi/\pi$. In the former case, as has already been noted (Table~\ref{t:01}), the kinematic parameters depend little on the constraint on the relative photometric distance error. In the latter case, the radius of the sample changes dramatically, which can affect the behavior of kinematic parameters. For example, the second derivative of the angular velocity of Galactic rotation, $\Omega^{''}_0$, is determined more confidently using stars farther from the Sun (Table~\ref{t:02}). Note that the solution obtained only from the line-of-sight velocities using the photometric distances is absent in Table~\ref{t:03} --- these results are presented in the middle part of Table~\ref{t:01}.

We think the solution obtained from the single OB stars with the photometric distances selected
under the condition $\sigma_r/r<10\%$ using the Gaia\,EDR3 proper motions (the first column in the
lower part of Table~\ref{t:03}) to be most interesting. In this solution we found the velocities $(U_\odot,V_\odot,W_\odot,)=(9.63,9.93,7.45)\pm(0.27,0.34,0.10)$~ km s$^{-1}$ and
 \begin{equation}  \label{sol-best-ERD3} \begin{array}{lll}
      \Omega_0 =~29.20\pm0.18~\hbox{km s$^{-1}$ kpc$^{-1}$},\\
  \Omega^{'}_0 =-4.150\pm0.046~\hbox{km s$^{-1}$ kpc$^{-2}$},\\
 \Omega^{''}_0 =~0.795\pm0.018~\hbox{km s$^{-1}$ kpc$^{-3}$},
 \end{array} \end{equation}
where the error per unit weight is $\sigma_0=9.56$~km s$^{-1}$ and $V_0=236.5\pm3.3$~km s$^{-1}$ (for the adopted distance $R_0=8.1\pm0.1$~kpc). This solution was obtained using (i) the original photometric distance estimates for the stars and (ii) the highly accurate proper motions of the OB stars. Note that all parameters of the solution~(\ref{sol-best-ERD3}) are in excellent agreement with the solution obtained virtually from the same stars using the Gaia\,EDR3 trigonometric parallaxes that is given in the first column in the lower part of Table~\ref{t:04}. At the same time, a smaller error per unit weight $\sigma_0$ was obtained in the solution~(\ref{sol-best-ERD3}).

\subsection*{Distance Scale Factor}
The ratio of the first derivative of the angular velocity of Galactic rotation found using only the proper motions to the one found only from the line-of-sight velocities is given in the last rows of Tables~\ref{t:01}--~\ref{t:04}. Following Zabolotskikh et al. (2002) and Rastorguev et al. (2017), we call this ratio the distance scale factor $p=(\Omega^{'}_0)_{V_l}/(\Omega^{'}_0)_{V_r}$. This is a correction factor of the form $p=r/r_{true}$, where $r$ are the distances being used and $r_{true}$ are the true distances, thus, $r_{true}=r/p$. We calculated the error of the factor $p$ based on the relation
$\sigma^2_p=(\sigma_{\Omega'_{0V_l}}/\Omega'_{0V_r})^2+
     (\Omega'_{0V_l}\cdot\sigma_{\Omega'_{0V_r}}/\Omega'^2_{0V_r})^2$.

The value of $p=0.86$ given in Table~\ref{t:01} implies that the photometric distances from Xiang et al. (2021)
should be lengthened by 14\%. The value of p = 0.86 is also confirmed by our analysis of the Gaia\,EDR3
proper motions (Table~\ref{t:03}). The value of $p=0.86$ given in Table~\ref{t:02}, which is close to unity, suggests that there is no need to lengthen the distance scale based on the Gaia\,DR2 trigonometric parallaxes. The values of $p=0.93-0.92$ given in Table~\ref{t:04} do not differ greatly from unity either.

 \begin{table}[p] \caption[]{\small
The kinematic parameters found from the OB stars based on all three conditional equations (\ref{EQ-1})--(\ref{EQ-3}) are given in the upper part of the table; those derived only from the Gaia\,EDR3 proper motions based on the two Eqs. (\ref{EQ-2}) and (\ref{EQ-3}) are given in the lower part of the table; the distances to the stars are photometric; $N_\star$ is the number of stars used, ${\overline r}$ is the mean radius of the sample.
 }
  \begin{center}  \label{t:03}
  \small
  \begin{tabular}{|l|r|r|r|r|r|}\hline
  Parameters &
   $\sigma_r/r<10\%$ & $\sigma_r/r<15\%$ & $\sigma_r/r<20\%$ & $\sigma_r/r<30\%$ \\\hline
  $N_\star$ &   9418 &              9628 &              9753 &            9793\\
  ${\overline r},$ kpc &       1.97 & 2.18 &         2.20 &  2.25\\
  &&&&\\
   $U_\odot,$ km s$^{-1}$ & $13.63\pm0.17$ & $13.87\pm0.16$ & $13.86\pm0.16$ & $13.96\pm0.16$\\
   $V_\odot,$ km s$^{-1}$ & $10.85\pm0.31$ & $11.50\pm0.30$ & $11.50\pm0.29$ & $11.52\pm0.29$\\
   $W_\odot,$ km s$^{-1}$ & $ 7.35\pm0.16$ & $ 6.95\pm0.14$ & $ 6.91\pm0.15$ & $ 6.94\pm0.15$\\
   $\Omega_0,$ km s$^{-1}$ kpc$^{-1}$
      &$ 31.05\pm0.17$ & $ 30.78\pm0.16$ & $30.75\pm0.16$ & $ 30.86\pm0.16$\\
  $\Omega^{'}_0,$ km s$^{-1}$ kpc$^{-2}$
      &$-4.349\pm0.043$& $-4.181\pm0.038$&$-4.169\pm0.038$& $-4.180\pm0.038$\\
 $\Omega^{''}_0,$ km s$^{-1}$ kpc$^{-3}$
      &$ 0.840\pm0.020$& $ 0.749\pm0.015$& $0.744\pm0.015$& $ 0.744\pm0.015$\\
   $\sigma_0,$ km s$^{-1}$ &       $14.76$ &       $15.03$ &       $15.16$ &       $15.32$\\
        $V_0,$ km s$^{-1}$ & $251.5\pm3.4$ & $249.3\pm3.3$ & $249.1\pm3.3$ & $250.0\pm3.4$\\
 \hline
   $U_\odot,$ km s$^{-1}$ & $9.63\pm0.27$ & $ 9.47\pm0.25$ & $ 9.40\pm0.25$ & $ 9.37\pm0.25$\\
   $V_\odot,$ km s$^{-1}$ & $9.93\pm0.34$ & $10.89\pm0.32$ & $11.00\pm0.32$ & $11.01\pm0.32$\\
   $W_\odot,$ km s$^{-1}$ & $7.45\pm0.10$ & $ 7.09\pm0.10$ & $ 7.05\pm0.10$ & $ 7.04\pm0.10$\\
   $\Omega_0,$ km s$^{-1}$ kpc$^{-1}$
      &$ 29.20\pm0.18$ & $ 28.91\pm0.17$ & $28.84\pm0.17$ & $ 28.80\pm0.17$\\
  $\Omega^{'}_0,$ km s$^{-1}$ kpc$^{-2}$
      &$-4.150\pm0.046$& $-3.965\pm0.040$&$-3.942\pm0.040$& $-3.936\pm0.040$\\
 $\Omega^{''}_0,$ km s$^{-1}$ kpc$^{-3}$
      &$ 0.795\pm0.018$& $ 0.711\pm0.014$& $0.703\pm0.014$& $ 0.702\pm0.014$\\
   $\sigma_0,$ km s$^{-1}$ &       $9.56$ &       $9.77$ &       $9.80$ &       $9.82$\\
        $V_0,$ km s$^{-1}$ & $236.5\pm3.3$ & $234.1\pm3.2$ & $233.6\pm3.2$ & $233.3\pm3.2$\\

 $(\Omega^{'}_0)_{V_l}/(\Omega^{'}_0)_{V_r}$
    & $0.86\pm0.01$ &  $0.86\pm0.01$ & $0.86\pm0.01$ &  $0.86\pm0.01$  \\ \hline
\end{tabular}\end{center} \end{table}

 \begin{table}[p] \caption[]{\small
The kinematic parameters found from the OB stars based on all three conditional equations (\ref{EQ-1})--(\ref{EQ-3}) are given in the upper part of the table; those derived only from the LAMOST line-of-sight velocities based on Eq.~(\ref{EQ-1}) are given in the middle part of the table; those derived only from the Gaia\,EDR3 proper motions based on the two Eqs.(\ref{EQ-2}) and (\ref{EQ-3}) are given in the lower part of the table; the distances to the stars were calculated based on the Gaia\,EDR3 parallaxes; $N_\star$ is the number of stars used, ${\overline r}$ is the mean radius of the sample.
 }
  \begin{center}  \label{t:04}
  \small
  \begin{tabular}{|l|r|r|r|r|r|}\hline
  Parameters &
     $\sigma_r/r<10\%$ & $\sigma_r/r<15\%$ & $\sigma_r/r<20\%$ & $\sigma_r/r<30\%$ \\\hline
    $N_\star$ &   9880 &             10545 &             10711 &           10821\\
  ${\overline r},$ kpc &       2.10 & 2.20 &              2.22 &            2.22\\
  &&&&\\
   $U_\odot,$ km s$^{-1}$ & $13.60\pm0.17$ & $13.72\pm0.17$ & $13.70\pm0.17$ & $13.73\pm0.16$\\
   $V_\odot,$ km s$^{-1}$ & $11.20\pm0.30$ & $11.24\pm0.30$ & $11.40\pm0.31$ & $11.31\pm0.31$\\
   $W_\odot,$ km s$^{-1}$ & $ 7.75\pm0.16$ & $ 7.65\pm0.15$ & $ 7.60\pm0.16$ & $ 7.58\pm0.15$\\

   $\Omega_0,$ km s$^{-1}$ kpc$^{-1}$
      &$ 30.22\pm0.17$ & $ 30.16\pm0.16$ & $30.07\pm0.16$ & $ 30.09\pm0.16$\\
  $\Omega^{'}_0,$ km s$^{-1}$ kpc$^{-2}$
      &$-4.160\pm0.040$& $-4.136\pm0.039$&$-4.106\pm0.039$& $-4.110\pm0.039$\\
 $\Omega^{''}_0,$ km s$^{-1}$ kpc$^{-3}$
      &$ 0.744\pm0.016$& $ 0.731\pm0.015$& $0.720\pm0.015$& $ 0.720\pm0.015$\\
   $\sigma_0,$ km s$^{-1}$ &       $15.20$ &       $15.45$ &       $15.43$ &       $15.43$\\
        $V_0,$ km s$^{-1}$ & $244.8\pm3.3$ & $244.3\pm3.3$ & $243.6\pm3.3$ & $243.7\pm3.3$\\
 \hline
   $U_\odot,$ km s$^{-1}$ & $14.58\pm0.28$ & $14.73\pm0.27$ & $14.78\pm0.27$ & $14.80\pm0.27$\\
   $V_\odot,$ km s$^{-1}$ & $11.65\pm0.58$ & $11.58\pm0.58$ & $11.65\pm0.58$ & $11.64\pm0.58$\\

  $\Omega^{'}_0,$ km s$^{-1}$ kpc$^{-2}$
      &$-4.431\pm0.096$& $-4.444\pm0.092$&$-4.439\pm0.092$& $-4.437\pm0.092$\\
 $\Omega^{''}_0,$ km s$^{-1}$ kpc$^{-3}$
      &$ 1.088\pm0.063$& $ 1.093\pm0.057$& $1.087\pm0.057$& $ 1.083\pm0.056$\\
   $\sigma_0,$ km s$^{-1}$ &      $21.00$ &       $22.39$ &       $22.48$ &       $22.47$\\
 \hline

   $U_\odot,$ km s$^{-1}$ & $9.67\pm0.28$ & $ 9.53\pm0.27$ & $ 9.37\pm0.27$ & $ 9.42\pm0.27$\\
   $V_\odot,$ km s$^{-1}$ & $9.84\pm0.34$ & $10.09\pm0.33$ & $10.13\pm0.33$ & $ 9.99\pm0.33$\\
   $W_\odot,$ km s$^{-1}$ & $7.89\pm0.11$ & $ 7.75\pm0.11$ & $ 7.72\pm0.11$ & $ 7.68\pm0.10$\\
   $\Omega_0,$ km s$^{-1}$ kpc$^{-1}$
      &$ 29.12\pm0.18$ & $ 29.02\pm0.17$ & $28.92\pm0.17$ & $ 28.97\pm0.17$\\
  $\Omega^{'}_0,$ km s$^{-1}$ kpc$^{-2}$
      &$-4.139\pm0.044$& $-4.085\pm0.041$&$-4.064\pm0.041$& $-4.079\pm0.041$\\
 $\Omega^{''}_0,$ km s$^{-1}$ kpc$^{-3}$
      &$ 0.756\pm0.016$& $ 0.732\pm0.014$& $0.726\pm0.014$& $ 0.729\pm0.014$\\
   $\sigma_0,$ km s$^{-1}$ &  10.22 & 10.36 &  10.40 &  10.45 \\
        $V_0,$ km s$^{-1}$ & $236.0\pm3.2$ & $235.1\pm3.2$ & $234.3\pm3.2$ & $234.6\pm3.2$\\

 $(\Omega^{'}_0)_{V_l}/(\Omega^{'}_0)_{V_r}$
    & $0.93\pm0.01$ &  $0.92\pm0.01$ & $0.92\pm0.01$ &  $0.92\pm0.01$  \\
 \hline
\end{tabular}\end{center} \end{table}

 \subsection*{Residual Velocity Ellipsoid for OB Stars}
As a result of the LSM solution of the conditional equations (\ref{EQsigm-1})--(\ref{EQsigm-6}) and our subsequent analysis, we determined the residual velocity dispersions of the OB stars calculated via the roots of the secular equation (\ref{ff-7}):
 \begin{equation}
 \begin{array}{lll}
  \sigma_1=18.29\pm0.27~\hbox{km s$^{-1}$},\\
  \sigma_2=11.75\pm0.29~\hbox{km s$^{-1}$},\\
  \sigma_3=~8.06\pm0.17~\hbox{km s$^{-1}$},
 \label{rezult-1}
 \end{array}
 \end{equation}
while the orientation of this ellipsoid is
 \begin{equation}
  \matrix {
  L_1=~83.7\pm1.5^\circ, & B_1=-1.0\pm0.6^\circ,\cr
  L_2=173.7\pm3.5^\circ, & B_2=-1.8\pm1.0^\circ,\cr
  L_3=144.2\pm1.5^\circ, & B_3=88.0\pm2.0^\circ.\cr
    }
 \label{rezult-11}
 \end{equation}
Obviously (see Fig.~\ref{f-10-XY}), the peculiarities of the sample, in particular, the absence of stars in the fourth quadrant, are responsible for the orientation of the first two axes. Interestingly, $B_3=88.0\pm2.0^\circ$, which, within the $1\sigma$ uncertainty, suggests the absence of significant ignored perturbations, since the third axis is essentially directed to the north Galactic pole. The photometric distances ($\sigma_r/r<10\%$), LAMOST line-of-sight velocities, and Gaia\,EDR3 proper motions were used here for the OB stars.

Applying the line-of-sight velocities increases the space velocity dispersion and leads to a biased estimate of the velocity $U_\odot$. We reanalyzed the residual velocities of the OB stars without using their line-of-sight velocities. As before, here we use the photometric distances of the OB stars and Gaia\,EDR3 proper motions. Now we seek the LSM solution of the system of only three conditional equations (\ref{EQsigm-1})--(\ref{EQsigm-3}). As a result, we found the following residual velocity dispersions of the OB stars:
 \begin{equation}
 \begin{array}{lll}
  \sigma_1=14.98\pm0.08~\hbox{km s$^{-1}$},\\
  \sigma_2=~8.86\pm0.05~\hbox{km s$^{-1}$},\\
  \sigma_3=~7.58\pm0.04~\hbox{km s$^{-1}$},
 \label{rezult-15}
 \end{array}
 \end{equation}
and the orientation parameters of this ellipsoid are
 \begin{equation}
  \matrix {
  L_1=~80.2\pm0.1^\circ,&B_1=+8.9\pm0.1^\circ,\cr
  L_2=170.2\pm0.1^\circ,&B_2=+0.1\pm0.1^\circ,\cr
  L_3=260.8\pm0.1^\circ,&B_3=81.1\pm0.1^\circ.\cr
    }
 \label{rezult-115}
 \end{equation}
We see that, in comparison with the results (\ref{rezult-1}) and (\ref{rezult-11}), the dispersions $\sigma_1,\sigma_2$, and $\sigma_3$ decreased, the errors in all parameters decreased significantly. In addition, here there are $B_1$ and $B_3$ that differ significantly from zero. In our opinion, in the solutions (\ref{rezult-1}) and (\ref{rezult-11}) the subtleties of the orientation of the velocity ellipsoid may be smeared out due to large errors in the parameters. The parameters (\ref{rezult-115}) show that the residual velocity ellipsoid for the OB stars is inclined to the Galactic plane. This inclination may be related to some perturbations in the vertical direction.
However, a nonuniform spatial distribution of stars most likely has an effect here (Fig.~\ref{f-10-XY}).
\section*{DISCUSSION}
The kinematics of OB stars with proper motions and parallaxes from the Gaia\,DR2 catalogue was analyzed
in Bobylev and Bajkova (2019).These OB stars were selected in Xu et al. (2018). Based on 5335
OB stars, with the correction $\Delta\pi=0.050$ mas having been added to their parallaxes, from the solution of only Eqs.~(\ref{EQ-2}) we found $(U_\odot,V_\odot)=(6.53,7.27)\pm(0.24,0.31)$~km s$^{-1}$ as well as $\Omega_0=29.70\pm0.11$~km s$^{-1}$ kpc$^{-1}$, $\Omega^{'}_0 =-4.035\pm0.031$~km s$^{-1}$ kpc$^{-2}$, and $\Omega^{''}_0 =~0.620\pm0.014$~km s$^{-1}$ kpc$^{-3}$, where the
error per unit weight is $\sigma_0=12.33$~km s$^{-1}$ and the linear rotation velocity of the Galaxy at the solar distance is $V_0=237.6\pm4.5$~km s$^{-1}$ (for the adopted $R_0=8.0\pm0.15$ kpc). A comparison of the corresponding parameters of this solution with those given in the lower part of Table~\ref{t:02} suggests their excellent agreement. This is not surprising, since the samples differ mainly by the method of applying the systematic correction to the Gaia\,DR2 trigonometric parallaxes.

Note the paper by Bobylev and Bajkova (2022), where based on the proper motions of 9720 OB stars,
we found the group velocity components $(U,V,W)_\odot=(7.21,7.46,8.52)\pm(0.13,0.20,0.10)$~km s$^{-1}$ and as well as
$\Omega_0 =29.712\pm0.062$~km s$^{-1}$ kpc$^{-1}$, $\Omega^{'}_0=-4.014\pm0.018$~km s$^{-1}$ kpc$^{-2}$, and $\Omega^{''}_0=0.674\pm0.009$~km s$^{-1}$ kpc$^{-3}$. Based on the proper motions of these OB stars, we determined the residual velocity dispersions $(\sigma_1,\sigma_2,\sigma_3)=(11.79,9.66,7.21)\pm(0.06,0.05,0.04)$~km s$^{-1}$ and showed that the first axis of this ellipsoid is slightly deflected from the direction toward the Galactic center, $L_1=12.4\pm0.1^\circ$, while the third axis is directed almost exactly to the north Galactic pole, $B_3=87.7\pm0.1^\circ$. We used a sample of OB2 stars from Xu et al. (2021) with proper motions and trigonometric parallaxes from the Gaia\,DR3 catalogue.

It is most interesting to note the solution found in Bobylev and Bajkova (2022) based on 1726 OB2 stars
with line-of-sight velocities and proper motions: $(U,V,W)_\odot=(7.17,10.03,8.15)\pm(0.30,0.35,0.29)$~km s$^{-1}$, $\Omega_0 =29.22\pm0.19$~km s$^{-1}$ kpc$^{-1}$, $\Omega^{'}_0 =-3.885\pm0.042$~km s$^{-1}$ kpc$^{-2}$, and $\Omega^{''}_0 =~0.685\pm0.031$~km s$^{-1}$ kpc$^{-3}$, where $\sigma_0=12.2$~km s$^{-1}$. Here we used the line-of-sight velocities that were copied by Xu et al. (2021) from the SIMBAD\footnote{http://simbad.u-strasbg.fr/simbad/} electronic database. When solving only Eq.~(\ref{EQ-1}), just as in this paper, the error per unit weight is about 25~km s$^{-1}$. A comparison of the parameters of this solution, for example, with those given in the first column in the upper part of Table~\ref{t:04} suggests that using the line-of-sight velocities from the LAMOST catalogue in the simultaneous solution leads to a bias of the $U_\odot$ estimate approximately by 6~km s$^{-1}$.

We lengthened the photometric distance scale. The results are presented in Fig.~\ref{f-dist-2}, where the distances to the OB stars calculated via the Gaia\,EDR3 trigonometric parallaxes are plotted against the photometric ones with different factors $p$. An analysis of the figure shows that the best agreement between the photometric distances and the Gaia\,EDR3 trigonometric scale is achieved when increasing the photometric distances by 10\% (Fig.~\ref{f-dist-2}a). Thus, we have fundamental agreement between our analysis of the proper motions and line-of-sight velocities (Tables~\ref{t:01} and \ref{t:03}) and the direct comparison of the distances (Fig.~\ref{f-dist-2}), which shows the necessity of lengthening the photometric distances from Xiang et al. (2021) approximately by 10\%.

When analyzing Fig.~\ref{f-dist-2}, we assumed, by default, that the distance scale of the Gaia\,EDR3 catalogue is exact. However, the Gaia\,EDR3 trigonometric parallaxes are known to have a slight offset $\Delta\pi$ relative to the inertial reference frame, which varies from $-0.015$ (Groenewegen 2021) to $-0.039$~mas (Zinn 2021). This correction should be added to the measured parallaxes and, therefore, the true distances to the stars must decrease. The correction depends strongly on the magnitude; it is not eliminated completely by simple methods.

\section*{CONCLUSIONS}
We studied the kinematics of a large sample of OB stars for which the photometric distances were
determined by Xiang et al. (2021) based on near-infrared data. We identified $\sim$15\,000 OB stars with the Gaia\,EDR3 catalogue. As a result, we analyzed the OB stars for which there are photometric distance estimates, line-of-sight velocities from the LAMOST catalogue, and trigonometric parallaxes and proper
motions from the Gaia\,DR2 and Gaia\,EDR3 catalogues. At the same time, our main sample contains
$\sim$10\,000 single OB stars with relative distance errors less than 10\%. In this paper we considered OB stars no farther than 5 kpc from the Sun, with a mean distance of about 2 kpc.

We showed that when seeking the kinematic parameters, using the line-of-sight velocities increases
significantly the space velocity dispersion. For example, this leads to a biased (by $\approx$3~km s$^{-1}$) estimate of the velocity $U_\odot$. In contrast, the mean residual velocity dispersion for the OB stars calculated only from their line-of-sight velocities is about 22~km s$^{-1}$, which is approximately twice its value found from their proper motions. Therefore, it is more advantageous to estimate the Galactic rotation parameters only from the Gaia proper motions.

We think the solution obtained from the single OB stars without using their line-of-sight velocities,
with photometric distances and proper motions from the Gaia\,EDR3 catalogue, to be most interesting. In
this solution we found the velocities $(U_\odot,V_\odot,W_\odot)=(9.63,9.93,7.45)\pm(0.27,0.34,0.10)$~km s$^{-1}$ as well as $\Omega_0=29.20\pm0.18$~km s$^{-1}$ kpc$^{-1}$, $\Omega^{'}_0=-4.150\pm0.046$~km s$^{-1}$ kpc$^{-2}$, and $\Omega^{''}_0=0.795\pm0.018$~km s$^{-1}$ kpc$^{-3}$, where the error per unit weight is $\sigma_0=9.56$~km s$^{-1}$ and $V_0=236.5\pm3.3$~km s$^{-1}$
(for the adopted distance $R_0=8.1\pm0.1$~kpc). The residual velocity ellipsoid for these OB stars has the following values of the principal axes: $(\sigma_1,\sigma_2,\sigma_3)=(15.13,9.69,7.98)\pm(0.07,0.05,0.04)$~km s$^{-1}$, the ellipsoid is inclined to the Galactic plane.

We considered three distance scales: (i) the photometric distances from Xiang et al. (2021), (ii) the
Gaia\,DR2 trigonometric parallaxes, and (iii) the Gaia\,EDR3 trigonometric parallaxes. The photometric
distances and the distances calculated based on the Gaia EDR3 parallaxes were shown to be in good
agreement up to distances of 4--5 kpc from the Sun. The various methods of analyzing the distance scales showed that the photometric distances from Xiang et al. (2021) should be lengthened approximately
by 10\%.

\section*{ACKNOWLEDGMENTS}
We are grateful to the referee for useful remarks that contributed to an improvement of the paper.

 \bigskip\medskip{REFERENCES}\medskip{\small
 \begin{enumerate}

 \item
V.V. Bobylev and A.T. Bajkova, Astron. Lett. {\bf 44}, 676 (2018). 

 \item
V.V. Bobylev and A.T. Bajkova, Astron. Lett. {\bf 45}, 331 (2019). 

 \item
V.V. Bobylev and A.T. Bajkova, Astron. Rep. {\bf 65}, 498 (2021). 

 \item
V.V. Bobylev and A.T. Bajkova, Astron. Rep. {\bf 66}, 269, (2022). 

 \item
R.L. Branham, Astrophys. J. {\bf 570}, 190 (2002).

 \item
R.L. Branham, Mon. Not. R. Astron. Soc. {\bf 370}, 1393 (2006).

 \item
A.G.A. Brown, A. Vallenari, T. Prusti, et al. (Gaia Collab.), Astron. Astrophys. {\bf 616}, 1 (2018). 

 \item
A.G.A. Brown, A. Vallenari, T. Prusti, et al. (Gaia Collab.), Astron. Astrophys. {\bf 649}, 1 (2021). 

 \item
J. Byl and M.W. Ovenden, Astrophys. J. {\bf 225}, 496 (1978).

 \item
B.-Q. Chen, et al., Mon. Not. R. Astron. Soc. {\bf 487}, 1400 (2019).

 \item
X.-Q. Cui, et al., Research Astron. Astrophys. {\bf 12}, 1197 (2012). 

 \item
A.K. Dambis, et al., Astron. Lett. {\bf 27}, 58 (2001).

 \item
D. Fern\'andez, et al., Astron. Astrophys. {\bf 372}, 833 (2001).

 \item
J.A. Frogel and R. Stothers, Astron. J. {\bf 82}, 890 (1977).

 \item
Y.M. Georgelin and Y.P. Georgelin, Astron. Astrophys. {\bf 49}, 57 (1976).

 \item
G.A. Gontcharov, Astron. Lett. {\bf 38}, 694 (2012).

 \item
M.A.T. Groenewegen, Astron. Astrophys. {\bf 654}, A20 (2021).

 \item
The HIPPARCOS and Tycho Catalogues, ESA SP--1200 (1997).

 \item
Z. Liu,et al., Astrophys. J. Suppl. Ser. {\bf 241}, 32 (2019).

 \item
J. Maiz Apell\'aniz, et al.,  Astrophys. J. Suppl. Ser. {\bf 151}, 103 (2004).

 \item
J. Maiz Apell\'aniz, et al.,  Astrophys. J. Suppl. Ser. {\bf 224}, 4 (2016).

 \item
A.M. Mel'nik and A.K. Dambis, Mon. Not. R. Astron. Soc. {\bf 400}, 518 (2009).

 \item
A.M. Mel'nik and A.K. Dambis, Mon. Not. R. Astron. Soc. {\bf 472}, 3887 (2017).

 \item
A.M. Melnik and A.K. Dambis, Astrophys. Space Science {\bf 365}, 112 (2020).

 \item
M. Miyamoto and Z. Zhu, Astron. J. {\bf 115}, 1483  (1998).

\item
K.F. Ogorodnikov, {\it Dynamics of stellar systems} (Oxford: Pergamon, ed. Beer, A. 1965).

\item
J.H. Oort, Bull. Astron. Inst. Netherland {\bf 3}, 275 (1927).

 \item
A.E. Piskunov, et al., Astron. Astrophys. {\bf 445}, 545 (2006).

 \item
M.E. Popova and A.V. Loktin, Astron. Lett. {\bf 31}, 663 (2005).

\item
T. Prusti, J.H.J. de Bruijne, A.G.A. Brown et al. (Gaia Collab.), Astron. Astrophys. {\bf 595}, 1 (2016). 

\item
A.S. Rastorguev, et al.,  Astrophys. Bulletin {\bf 72}, 122 (2017).

\item
B.C. Reed, PASP {\bf 107}, 907 (1995).

 \item
D. Russeil, Astron. Astrophys. {\bf 397}, 133 (2003).

 \item
B.A. Skiff, VizieR Online Data Catalog, B/mk (2014).

 \item
A. Sota, et al., Astrophys. J. Suppl. Ser. {\bf 193}, 24 (2011).

 \item
N. Tetzlaff, et al., Mon. Not. R. Astron. Soc.  {\bf 410}, 190 (2011).

 \item
J. Torra, et al., Astron. Astrophys. {\bf 359}, 82 (2000).

\item
M. Uemura, et al., Publ. Astron. Soc. Japan  {\bf 52}, 143 (2000).

 \item
W. Wegner, Mon. Not. R. Astron. Soc. {\bf 319}, 771  (2000).

 \item
M.-S. Xiang, et al., Mon. Not. R. Astron. Soc. {\bf 467}, 1890 (2017). 

 \item
M. Xiang, et al., Astrophys. J. Suppl. Ser.{\bf 253}, 22 (2021).

 \item
Y. Xu, et al., Astron. Astrophys. {\bf 616}, L15 (2018).

 \item
Y. Xu, et al., Astron. Astrophys. {\bf 645}, L8 (2021).

 \item
P.A. Zaal, et al.,  Astron. Astrophys. {\bf 366}, 241 (2001).

 \item
M.V. Zabolotskikh, et al.,  Astron. Lett. 28, 454 (2002).

 \item
P.T. de Zeeuw, et al., Astron. J. {\bf 117}, 354 (1999).

 \item
Z. Zhu, Chin. J. Astron. Astrophys. {\bf 6}, 363 (2006).

 \item
J.C. Zinn, Astron. J. {\bf 161}, 214 (2021).

 \end{enumerate} }
\end{document}